\input harvmac
\noblackbox
%
%

\font\tiau=cmcsc10
\baselineskip 12pt
\Title{\vbox{\baselineskip12pt \hbox{}
\hbox{} }}
{\vbox{\hbox{\centerline{\bf S-DUALITY, SL(2,Z) MULTIPLETS AND KILLING SPINORS}}}}
\vskip3cm
\centerline{\tiau Pablo M. Llatas\foot{llatas@fpaxp1.usc.es}}
\centerline{\tiau Jose M. S\'anchez de Santos\foot{santos@gaes.usc.es}}

\vskip.1in
\centerline{\it Departamento de F\'\i sica de Part\'\i culas}
\centerline{\it Universidad de Santiago}
\centerline{\it Santiago de Compostela, E-15706}
\vskip .9cm
\centerline{\bf Abstract}
\vskip .9cm
\noindent{The S-duality transformations in type IIB string theory can be
seen as local $U(1)$ transformations in type IIB supergravity. We use
this approach to construct the $SL(2,Z)$ multiplets associated to 
supersymmetric backgrounds of type IIB string theory and 
the transformation laws of their corresponding Killing spinors. 
}

\vskip2cm
US-FT-24/99

hep-th/9912159

\Date{December 99}

\lref\sw{J.H. Schwarz, Nucl. Phys. B226(1983) 269-288.}

\lref\dW{B. de Wit, D.J. Smit and N.D. Hari Dass, Nucl. Phys. B283(1987) 165-191.}

\lref\ib{A. Font, L. Iba\~ nez, D. Lust and F. Quevedo, Phys. Lett. B249 (1990) 35.}

\lref\lr{J.X. Lu and S. Roy, Phys. Lett. B428(1998)289, hep-th/9802080.}

\lref\schw{J. H. Schwarz, Phys. Lett. B360(1995)13; 
Erratum-ibid. B364(1995)252, hep-th/9508143.}

\lref\o{T. Ort\'\i n, Phys. Rev. D51(1995)790, hep-th/9404035.}

\lref\im{Y. Imamura, Nucl. Phys. B537(1999) 184, hep-th/9807179.}

\lref\dab{A. Dabholkar, G. Gibbons, J.A. Harvey and F. Ruiz Ruiz, Nucl. Phys.
B 340 (1990) 33.}

\lref\has{S.F. Hassan, hep-th/9907152.}

\lref\hor{G. Horowitz and A. Strominger, Nucl. Phys. B360(1991) 197.}

\lref\mal{Juan M. Maldacena, Ph. D Thesis, hep-th 9607235.}

\lref\mp{R.C. Myers and M.J. Perry, Ann. Phys. 172 (1986) 304.}

\lref\hyt{C. Hull and P. Townsend, Nucl. Phys. B438 (1995) 109, hep-th/9410167.}

\lref\gibb{G.W. Gibbons, M.B. Green and M.J. Perry, 
Phys. Lett. B370(1996)37, hep-th/9511080.}

\lref\dsi{A. Dabholkar, ``Lectures on Orientifolds" hep-th/9804208
and references
therein. P. Meessen and T. Ort\'\i n, Nuc. Phys. B541(1999)195, hep-th/9806120. E. Eyras
and Y. Lozano, hep-th/9908094.}

\lref\witten{E. Witten, Nucl. Phys. B460(1996)335, hep-th/9510135.}

\lref\pop{H. Lu, C.M. Pope and J.Rahmfeld, hep-th/9805151.}

\baselineskip 12pt

\newsec{Introduction}

The low energy limit of type IIB superstring theory is type IIB supergravity
\refs{\sw ,\dW}. Among the symmetries of type IIB string theory, S-duality 
is peculiar and
has concentrated a lot of work since its discovery \refs{\ib}, since it
is a non-perturbative (conjectured quantum exact) symmetry 
relating the strong and weak coupling regimes of string theory.
This S-duality symmetry is encoded, classically, in a $SL(2,R)$ group,
which is broken in the quantum theory to $SL(2,Z)$. The existence of
this symmetry ensures that any solution of type IIB
string theory belongs to a $SL(2,Z)$ multiplet. 

The
$SL(2,Z)$ multiplets to which the fundamental string and the 
solitonic five branes belong have been identified in \refs{\schw} and
\refs{\lr} respectively. In this paper we will study a more general case.
Our expressions will apply for any solution of the type IIB 
string effective equations of motion whith at least one non 
vanishing electric or magnetic charge with respect to the 
three forms (or their duals) appearing in the theory.
We identify the $SL(2,Z)$ multiplet to which a given type IIB string solution
of this type
belongs. We will also study the behaviour of the Killing spinors associated
to supersymmetric solutions under S-duality transformations. This is particularly
interesting when studying the world-volume theories of extended objects in IIB
backgrounds, since it allows to write the conditions for BPS solutions coming from
kappa symmetry.

This paper is organized as follows. First we review quickly 
the aspects of type IIB supergravity and type IIB superstring theory
that will be used through the paper, to fix our conventions. The equations of motion
of type IIB supergravity 
are invariant under a local $U(1)$ transformation. Second, we introduce a
map relating the equations of motion of type IIB supergravity
and type IIB superstring theory and recognize the 
local $U(1)$ invariance of type IIB supergravity as S-duality
transformations of type IIB string theory. This allows one to identify
the $SL(2,Z)$ multiplets associated to a given solution of type IIB
string theory, and provide us automatically with
the transformation law of the Killing spinors of the theory under
S-duality. Then we work
out explicitly the cases of magnetically and electrically 
charged $SL(2,Z)$ multiplets, checking our general expressions with 
examples previously constructed in the literature. 
In the last section we present our
conclusions and discussion.

\newsec{Quick Review of Type IIB Supergravity and the Type IIB String.}

The equations of motion for ten dimensional chiral $N=2$ 
supergravity have been known for a long time (\refs{\sw ,\dW}). 
They are given by (for the bosonic part of the
theory):

\eqn\dW{
\eqalign{
&D^{\mu}P_{\mu}=-{1\over 24}G^{\mu\nu\rho}G_{\mu\nu\rho} \cr
&D^{\rho}G_{\mu\nu\rho}=P^{\rho}G^{*}_{\mu\nu\rho} -{2\over 3} i 
  K_{\mu\nu\rho\lambda\sigma}G^{\rho\lambda\sigma}\cr
&R_{\mu\nu}=P_{\mu}P^{*}_{\nu}+P_{\nu}P^{*}_{\mu}+
           {1\over 8}(G^{*\alpha\beta}_{\mu}G_{\nu\alpha\beta}+
                      G_{\mu}^{\alpha\beta}G^{*}_{\nu\alpha\beta})-
           {1\over 6}g_{\mu\nu}G^{*}_{\tau\alpha\beta}G^{\tau\alpha\beta}
+{1\over 6} K_{\alpha\beta\gamma\sigma\mu}K^{\alpha\beta\gamma\sigma}_{\nu}\cr 
& K=*K\cr
}}
where $D_{\mu}$ stands for the covariant derivative with respect to the $U(1)$
gauge field $Q_{\mu}$ present in type IIB supergravity 
($D_{\mu}=\nabla_{\mu}-iq Q_{\mu}$, with $q=2,1$
for $P_{\mu}$ and $G_{\mu\nu\rho}$ respectively) and $*$ denotes Hodge duality. 

These equations have well known invariances. One of them (easy to check from
\dW ) is the local $U(1)$ transformation:
\eqn\u{
\eqalign{
&P\to P' =e^{2 i \Lambda (x)} P\cr
&G\to G' =e^{ i \Lambda (x)} G\cr
&Q\to Q' =Q +d\Lambda\cr
&K\to K' =K\cr
&g_{\mu\nu}\to g'_{\mu\nu} =g_{\mu\nu}\cr
}}

We could consider solutions to the equations \dW\ with
excitations of the five form $K$. 
But, as we can see in \u ,
these excitations are neutral under $U(1)$ transformations and, for this
reason, will not enter in the discussions below. We set this five form
to zero from now on.

A second symmetry of the equations \dW\ are chiral $N=2$ supersymmetryc
tranformations. In particular, the ``dilatino" and ``gravitino" 
transformations under supersymmetry are given by:

\eqn\dWs{
\eqalign{
& \delta\lambda =i P_{\mu}{\Gamma}^{\mu}{\epsilon}^{*} -{i\over 24}G_{\mu\nu\rho}
        {\Gamma}^{\mu\nu\rho}{\epsilon} \cr
& \delta\Psi_{\mu} =D_{\mu}\epsilon +{1\over 96}
     (G_{\nu\alpha\beta}\Gamma_{\mu}^{\,\,\,\nu\alpha\beta}
-9 G_{\mu\alpha\beta}\Gamma^{\alpha\beta}){\epsilon}^{*}.\cr
}}
In the equation above, $\lambda$ and $\Psi_\mu$ are complex Weyl spinors of opposite chirality,
$\Gamma_{11}\lambda=\lambda$, $\Gamma_{11}\Psi_\mu=-\Psi_\mu$. The supersymmetry parameters
$\epsilon$ form also a complex Weyl spinor satisfying $\Gamma_{11}\epsilon=-\epsilon$.
Notice that these supersymmetric transformations are covariant under the local $U(1)$
transformations of \u\ (the charges of $\lambda$, $\Psi_{\mu}$ and
$\epsilon$ being $3/2$, $1/2$ and $1/2$ respectively). In particular, 
BPS states will remain BPS 
after a local $U(1)$ transformation.

The bosonic sector of the low energy limit of
type IIB string theory is described by the action (when the five form is turned
off and in the Einstein frame)
\eqn\S{
S_{IIB}={1\over {\alpha '}^4}
\int{d^{10}x\sqrt{-g} [ R-{1\over 2}\partial_{\mu}\Phi\partial^{\mu}\Phi-
{1\over 2} e^{2\Phi}\partial_{\mu}\chi\partial^{\mu}\chi-
{1\over 12}(e^{-\Phi}H^{(1)2}+e^{\Phi}F^2)] }
}
where $\alpha '=l_s^2$ is the string tension and we have defined
\eqn\dej{
\eqalign{
&H^{(1)}=dB,\cr
&F=H^{(2)}-\chi H^{(1)}=dC-\chi dB
}}
(being $B$ and $C$ the NS-NS and R-R 2-form potentials respectively).

\newsec{IIB String S-Duality as Local $U(1)$ IIB Supergravity Transformations.}

The equations of motion of type IIB supergravity \dW\ exactly
map to the equations of motion derived from the IIB superstring action \S\ if we
make the following identifications between the string and 
supergravity fields (we use the language of forms):

\eqn\id{
\eqalign{
&P={1\over 2}(d\Phi+i e^{\Phi}d\chi)\cr
&Q=-{1\over 2} e^{\Phi}d\chi\cr
&G=e^{-{1\over 2}\Phi}H^{(1)}+i e^{{1\over 2}\Phi}
   F .\cr
}}
This map corresponds to a particular parametrization and gauge
choice for the supergravity scalar fields and was also obtained in \refs\has . 
It can be partially inverted to obtain:
\eqn\in{
\eqalign{
&d\Phi =2(P+i Q)\cr
&d\chi =-2 Q e^{-\Phi} \cr
&H^{(1)}= e^{{1\over 2}\Phi} {\rm Re} G\cr
 &H^{(2)}= e^{-{1\over 2}\Phi} {\rm Im} G+ e^{{1\over 2}\Phi} \chi{\rm Re} G, \cr
}}
where ${\rm Re}G$ and ${\rm Im}G$ denote the real and imaginary parts of $G$.
The action \S\ 
can then be written as:
\eqn\pab{
S_{IIB}=\int{d^{10}x\sqrt{-g} [R-2 \| P{\|}^2 -{1\over 12}\| G{\|}^2]},
}
where the $U(1)$ invariance \u\ is obvious.

Now, using \u\ it is straightforward to obtain from \in\ the  
local $U(1)$ transformation laws
of $d\Phi$, $d\chi$, $H^{(1)}$ and $H^{(2)}$. If we demand $d\Phi '$ to be a real
field, this imposses an additional condition on the allowed 
local $U(1)$ transformations.
$\Lambda(x)$ has to satisfy the equation:
\eqn\la{
d\Lambda(x)=-\sin(\Lambda(x)) \cos(\Lambda(x))d\Phi +\sin^2 (\Lambda(x))e^{\Phi}
d\chi,
}
whose solution is:

\eqn\laa{
\Lambda (x)=- \arctan({k e^{-\Phi (x)}\over {1+k\chi (x)}}),
}
being $k$ a constant of integration.

Under local $U(1)$ transformations with this $\Lambda (x)$, 
one gets from \in\ the transformation laws of $d\Phi$ and $d\chi$:

\eqn\dd{
\eqalign{
d\Phi' &=\cos(2\Lambda (x)) d\Phi- \sin(2\Lambda (x)) e^{\Phi} d\chi\cr
e^{\Phi '}d\chi '&=\sin(2\Lambda (x)) d\Phi+\cos(2\Lambda (x))e^{\Phi} d\chi .\cr
}}
These equations can be easily integrated and one obtains:
\eqn\da{
\eqalign{
e^{\Phi '}&= e^{\Phi} [c^2 ({\chi}^2+e^{-2 \Phi})+2 c d\chi +d^2]\cr
\chi ' &={{a c ({\chi}^2+e^{-2 \Phi})+(a d +b c)\chi + b d}\over 
    { [c^2 ({\chi}^2+e^{-2 \Phi})+2 c d\chi +d^2]}}\cr
}}

Now, from \in\ one can also compute the transformation laws for $H^{(1)}$ and $H^{(2)}$.
One gets:
\eqn\h{
\eqalign{
H^{(1) '}&=d H^{(1)}+ c H^{(2)}\cr
H^{(2) '}&=b H^{(1)}+ a H^{(2)}.\cr
}}
Note that \da\ and \h\ are nothing but the usual S-duality transformations 
of type IIB superstring theory realized as
$U(1)$ local transformations in type IIB supergravity. To make contact
with the usual $SL(2,R)$ language of S-duality we have just
redefined the two integration constants coming from \dd\ together with $k$ in \laa\
into the four constants $a$, $b$, $c$
and $d$ plus the $SL(2,R)$ constraint $ad-bc=1$.
After this redefinition, $k=c/d$
and the local $U(1)$ gauge transformation
function $\Lambda (x)$ can be determined through its sine and cosine
:

\eqn\y{
\sin \Lambda (x)=-{{c e^{-\Phi}}\over{\sqrt{(d+c\chi)^2+c^2 e^{-2\Phi}}}}, 
\,\,\,\,\,\,\,\,\,\,\,\,\,\,
\cos \Lambda (x)={{d+c\chi}\over{\sqrt{(d+c\chi)^2+c^2 e^{-2\Phi}}}}.
}

This automatically gives the transformation law of the Killing spinors under 
S-duality transformations. The Killing spinors $\epsilon$ have ${1\over 2}$
$U(1)$ charge, therefore:
\eqn\ks{
\epsilon '(x)= e^{{i\over 2}\Lambda (x)}\epsilon (x),
}
with $\Lambda (x)$ given in \y \foot{Similar expressions were derived 
in \refs{\o ,\im} for specific cases.}.

The transformation of the Killing spinors can be understood (formally) as a parallel 
transport using the $U(1)$ connections $Q$ and $Q'$ as follows. Let us take 
a ten dimensional space-time point $x_0$ which is a fixed point of the
local $U(1)$ transformation (i.e.,
$\Lambda (x_0)=0$). Then, from \u , we have that
\eqn\pt{
\Lambda (x)=\int^{x}_{x_0}d\Lambda =\int^{x}_{x_0} (Q' -Q)=
 \int^{x}_{x_0} Q' +\int^{x_0}_{x} Q.
}
As a result (see \ks ), the effect of a S-duality transformation on a spinor
can be understood as a parallel transport of the spinor $\epsilon (x)$ from
any point $x$ of space-time to a fixed point $x_0$ of the $U(1)$ 
transformation with the $Q$ connection, and then parallel transport
the resulting spinor back to the initial point $x$ with the gauge connection $Q'$.
If we perform a $SL(2,R)$ transformation with $c=0$, any point in space-time
is a fixed $U(1)$ point. We can then take $x_0 =x$ in \pt\ and the Killing
spinor remains invariant. 

There can be also fixed $U(1)$ points 
where $\Phi=\infty$. These points are singular and the classical solutions
to the effective string equations are not reliable (there are 
strong quantum gravity stringy corrections in a neighborhood of these points
of size $\sim l_s$).
So the parallel transport \pt\ starts
from a point where the classical solution of the 
effective string action
is reliable, goes to a singular point, 
where quantum gravity effects take place, and comes back to the initial
point where  the classical solution of the 
effective string action
is reliable again (note that $\Lambda (x)$ is well defined all along
the path). 
In this way the S-duality transformations capture non-perturbative
information from the supergravity point of view.

\newsec{Magnetic Multiplets.}

Given any solution of the equations of motion 
$g_{\mu \nu}$, $\Phi$, $\chi$, $B$ and
$C$, all the members of the $SL(2,Z)$ multiplet associated to it  can be obtained
from \da\ and \h\ 
once the right parameters  $a$, $b$, $c$ and $d$ connecting them are correctly
identified. This is what we will be doing in this (next) section, distinguishing
the cases in which at least one of the magnetic (electric) charges are non-vanishing.  

Let us first consider the simplest case in which 
there is at least one non-vanishing 
magnetic charge in the solution to the equations of motion.
Let us define the magnetic 
charges in the following way:

\eqn\ch{
q_i =\int{H^{(i)}},\,\,\,\,\,\,\,\,\,\,\,p_i =\int{H^{(i)'}},
}
where the integration is performed over a 
region that surrounds the world-volume of the charged object.
One can use the four equations \da\ and \h\ to obtain the three 
independent values of the 
constants $a$, $c$ and $d$ ($b$ is obtained from $ad-bc=1$) 
in terms of those charges and the values
$\Phi_0$, $\chi_0$, $\Phi_0 '$ and $\chi_0 '$ of the fields 
$\Phi$, $\chi$, $\Phi '$ and $\chi '$ evaluated at some point (which
is usually taken at infinity). The result is the following:

\eqn\cons{
\eqalign{
a &={1\over\Delta}[{q_1}^2 (e^{\Phi_0 '}\chi_0 '+e^{\Phi_0}\chi_0)-
   e^{\Phi_0}(p_1 p_2+q_1 q_2)]\cr
b &={1\over\Delta}[(p_1 p_2 -q_1q_2)(e^{\Phi_0 '}\chi_0 '+e^{\Phi_0}\chi_0)+
   (e^{\Phi_0}q_2^2-e^{\Phi'_0}p_2^2)]\cr
c &={1\over\Delta}({q_1}^2 e^{\Phi_0 '}- {p_1}^2 e^{\Phi_0})\cr
d &={1\over\Delta}[{p_1}^2 (e^{\Phi_0 '}\chi_0 '+e^{\Phi_0}\chi_0)-
   e^{\Phi_0 '}(p_1 p_2+q_1 q_2)],\cr
}}
being $\Delta$ the constant:
\eqn\del{
\Delta \equiv q_1 p_1 (e^{\Phi_0 '}\chi_0 '+e^{\Phi_0}\chi_0)-
 (q_1 p_2 e^{\Phi_0 '}+ q_2 p_1 e^{\Phi_0}).
}
Note that, since we are assuming that  $q_1$ and/or $q_2$ do not vanish, 
the constants \cons\ are generically well defined. It is clear from \cons\ that
quantization conditions on the charges $q_1 , q_2 , p_1 , p_2$
break $SL(2,R)$ down to $SL(2,Z)$ \refs\hyt\ (for fixed asympotic values of the
scalar fields).Equation \cons\ provides the $SL(2,Z)$ parameters
relating any two components of the multiplet.
Since we started from four equations (\da\ and \h ) and three unknowns
($a$, $c$ and $d$), we also get 
a constraint equation that $q_i$, $p_i$,
$\Phi$, $\chi$, $\Phi '$ and $\chi '$  have to satisfy:
\eqn\conss{
e^{\Phi} (q_2 - q_1 \chi )^2+e^{-\Phi} {q_1}^2 =
e^{\Phi '} (p_2 - p_1 \chi ')^2+e^{-\Phi '} {p_1}^2.
}
The interpretation of this constraint will be clear in what follows. The
Killing spinors are straightforwardly computed
by using the equations \ks , \y\ and 
\cons .

Lets us work out a specific example. We will take a BPS NS5-brane 
in the Einstein frame:

\eqn\five{
\eqalign{
dS^2{_E}&=H(r)^{-{1\over 4}}(-dt^2 +dX^2_{\|})+H(r)^{3\over 4}
(dr^2+r^2 d\Omega_3)\cr
e^{\Phi}&=\sqrt{H(r)},\,\,\,\,\,\,\,\,\,\,\, H^{(1)}=2 R^2 \epsilon_3,
 \,\,\,\,\,\,\,\,\,\,H^{(2)}=\chi =0,\cr
}}
being $H(r)$ the harmonic function:
\eqn\ar{
H(r)=1+{R^2\over r^2},
}
and construct the $SL(2,Z)$ multiplet to which it belongs,
and the corresponding Killing spinors (this
multiplet was first constructed in \refs\lr ).

We will take, moreover, the 
asymptotic conditions, $\Phi_{0}'\equiv \Phi '(\infty )$
and $\chi_{0}'\equiv \chi '(\infty )$
(note that in \ar\ we have already taken $\Phi_{0}\equiv \Phi (\infty )=0$
and $\chi_{0}\equiv \chi (\infty )=0$).

From equation \conss\ we obtain, for this case: 
\eqn\cn{
q{_1}^2=e^{\Phi_0 '} (p_2 - p_1 \chi_0 ')^2+e^{-\Phi_0 '} {p_1}^2
}
which allows to eliminate $q_1$ in equations \cons\
and write them in terms of $p_1$ and $p_2$ alone. This gives:
\eqn\ro{
\eqalign{
a &={1\over q_1} [e^{\Phi_0 '}\chi_0 '(p_1 \chi_0 ' - p_2)+p_1 e^{-\Phi_0 '}]\cr
b &={p_2\over q_1} \cr
c &={1\over q_1} e^{\Phi_0 '} (p_1 \chi_0 ' - p_2)\cr
d &={p_1 \over q_1} .\cr
}}
Just plugging this result
on \da\ and \h\ we get the $SL(2,Z)$ multiplet:
\eqn\mul{
\eqalign{
e^{\Phi '}&=c^2 H(r)^{-{1\over 2}}+d^2 H(r)^{1\over 2}\cr
\chi ' &={{ac +bdH(r)}\over {c^2 +d^2 H(r)}}\cr
H^{(1)'} &=d H^{(1)}\cr
H^{(2)'} &=b H^{(1)}.\cr
}}

The Einstein metric is inert under the $U(1)$ transformations
(see equation \u ), meaning that the mass 
measured with it is equal for all the  members of the multiplet. The
coefficient $R$ of the harmonic form is then invariant. Using \cn, it can be 
written in terms of the charges $p_1$, $p_2$:
\eqn\ta{
q_1 =\int{H^{(1)}}=2 R^2 w_3\Rightarrow R_{(p_1,p_2)}^2\equiv R^2=
{\sqrt{{e^{\Phi_0 '} (p_2 - p_1 \chi_0 ')^2+e^{-\Phi_0 '} {p_1}^2}}\over{2 w_3}},
}
where we denoted by $w_3$ the volume of the unit $S^3$.
Usually, the charges are given in units in which $q_1=1$. The NS$5$-brane and
the D$5$-brane are then just the $(1,0)$ and $(0,1)$ elements of this multiplet
of $(p_1, p_2)$ magnetic 5-branes.

Let us compute the Killing spinors. 
In the present case, the function $\Lambda$ depends only on $r$ and
can be written in terms of the charges $p_1$ and $p_2$ from the equations \ro\ as:
\eqn\lam{
\eqalign{
\sin \Lambda_{(p_1,p_2)}(r) & =
-{{e^{\Phi'_0} (p_1 \chi'_0-p_2)}\over
\sqrt{e^{2 \Phi'_0} (p_1 \chi'_0-p_2)^2 + p_1^2 H(r)}}\cr
\cos \Lambda_{(p_1,p_2)}(r) & =
{{p_1 [H(r)]^{1\over 2}}\over
\sqrt{e^{2 \Phi'_0} (p_1 \chi'_0-p_2)^2 + p_1^2 H(r)}}.\cr
}}
Now, using equation \ks , the Killing spinor of the $(p_1,p_2)$ magnetic 5-brane
can be obtained from that of the NS5-brane as:
\eqn\kspq{
\epsilon_{(p_1,p_2)}=
e^{{i\over 2}\Lambda_{(p_1,p_2)(r)}}\epsilon_{NS5},
}
where
\eqn\killns{
\epsilon_{NS5}=
\left[ H(r)\right]^{-{1\over {16}}}
e^{-{\theta^3 \over 2} \Gamma_{\underline{\theta^3 r}}} 
\,\,e^{-{\theta^2 \over 2} \Gamma_{\underline{\theta^2 \theta^3}}}
\,\,e^{-{\theta^1 \over 2} \Gamma_{\underline{\theta^1 \theta^2}}}
\,\,\epsilon_0
}
being $\epsilon_0$ a constant complex Weyl spinor of negative chirality satisfying:
\eqn\pro{
\Gamma_{\underline{012345}} \epsilon_0=\epsilon_0^*.
}
In eq. \killns\ the $\Gamma$'s with underlined indices represent flat (constant) 
ten dimensional Dirac gamma matrices and $\theta^1,\theta^2,\theta^3$ are 
the angular coordinates of the transverse 3-sphere. Notice that the angular 
dependent part in \killns\ is just the Killing spinor of $S^3$ 
(\refs{\pop}).

We have already seen that the ADM masses, computed with the Einstein metric,
are equal for all members of the multiplet. However,  
if we want to visualize the
kind of objects that belong to the multiplets from the string theory
point of view (i.e., the dependence of the masses on the string
coupling constant), we have to compute the ADM masses with 
the modified Einstein metric 
${\tilde g}_{\mu\nu}$\refs{\mal}. This metric is related to the Einstein metric by:

\eqn\tg{
{\tilde g}_{\mu\nu}\equiv\sqrt{g_s} g_{\mu\nu}=e^{{\Phi_0 '}\over 2}g_{\mu\nu}.
}
The modified Einstein metric is defined such that it coincides with the string metric
asymtotically. It is clear from \tg\ that the masses of the members of the
multiplet, measured with ${\tilde g}_{\mu\nu}$, are different (due to
the change of the string coupling contant under $SL(2,Z)$ transformations).
If we write the action \S\ in terms of the modified Einstein metric,
the Newton constant $G_N$ gets rescaled to $G_N=g_s^2 {\alpha '}^4$. The ADM
mass is now computed from the behaviour of ${\tilde g}_{00}$ at 
infinity \refs\mp:
\eqn\adm{
{\tilde g}_{00}\sim -1+{G_N\over {3 w_3}} {M\over r^2}.
}
In our case, after writing the modified 
Einstein metric associated to the metric in equation \five , we get:
\eqn\oab{
{\tilde g}_{00}\sim -1+{1\over 4} {{\sqrt{g_s} R_{\bar p}^2}\over r^2}
}
Comparing \adm\ and \oab , and using the equation \ta , we arrive at:
\eqn\guay{
M={3\over {8{\alpha '}^4 }}\sqrt{g_s^{-2}(p_2-p_1\chi_0 ')^2+g_s^{-4}p_1^2}.
}
Note that this ADM mass has units of mass per unit of brane volume ($1/(Lenght^6)$),
so it corresponds to the tensions of the $(p_1,p_2)$ five branes. Note also
that the $p_1$ charged objects ``contribute" to the mass with terms of order
$p_1/g_s^2$ (as corresponds to the magnetic NS five brane) whereas
the $p_2$ charged objects ``contribute" to the mass with a term of order
$p_2/g_s$ (as corresponds to D-branes solitons). Then, \mul\ describes
the background configurations corresponding to bound states of
solitonic 5-branes and D5-branes \refs{\witten}.

\newsec{The Electric Multiplets.}

Let us now consider the cases in which there is at least one non-vanishing
electric charge. In the electric case, 
the conserved charges are given by the equations of motion for 
the NS-NS and R-R potentials $B$ and
$C$:
\eqn\om{
d*S^{(1)}=d*S^{(2)}=0,
}
where we have defined
\eqn\ss{
\eqalign{
S^{(1)}&\equiv e^{-\Phi} H^{(1)}-e^{\Phi}\chi F\cr
S^{(2)}&\equiv e^{\Phi} F.\cr
}}
Then, we define the electric charges as:

\eqn\chu{
{\hat q}_i =\int{*S^{(i)}},\,\,\,\,\,\,\,\,\,\,\,{\hat p}_i =\int{*S^{(i)'}},
}
where the integration region surrounds the charged object.

From the equations \da\ and \h\ we can work out the transformations laws for
$S^{(1)}$ and $S^{(2)}$ under local $U(1)$ transformations. The result 
is the following:
\eqn\ge{
\eqalign{
S^{(1) '}&=a S^{(1)}-b S^{(2)}\cr
S^{(2) '}&=-cS^{(1)}+d S^{(2)},\cr
}}
i.e., using \chu: 
\eqn\gus{
\eqalign{
{\hat p}_1 &=a {\hat q}_1 -b {\hat q}_2\cr
{\hat p}_2 &=-c {\hat q}_1 +d {\hat q}_2.\cr
}}
Note that, compared with \h , if the magnetic charges transforms
with the $SL(2,Z)$ matrix $K$, then the electric charges transforms
with $(K^{-1})^T$. Comparing \ge\ with \h\ we see that,
in the electric case, we have to make the changes $q_1\to 
{\hat q}_2$,
$q_2\to -{\hat q}_1$, $p_1\to {\hat p}_2$ and $p_2\to -{\hat p}_1$ in the
equations \cons ,\del and \conss\foot{This is nothing but the electric-magnetic
duality $H^{(1)}\to *S^{(2)}$, $H^{(2)}\to - *S^{(1)}$.}. 
These changes give, for the 
$SL(2,Z)$ constants, the expressions:

\eqn\col{
\eqalign{
a &={1\over{\hat{\Delta}}}[{{\hat q}_2}^2 (e^{\Phi_0 '}\chi_0 '+e^{\Phi_0}\chi_0)+
   e^{\Phi_0}({\hat p}_1 {\hat p}_2+{\hat q}_1 {\hat q}_2)]\cr
b &={1\over {\hat{\Delta}}}[({{\hat q}_1} {{\hat q}_2}-{{\hat p}_1}{{\hat p}_2})
(e^{\Phi_0 '}\chi_0 '+e^{\Phi_0}\chi_0)+
({{\hat q}_1}^2 e^{\Phi_0 }- {{\hat p}_1}^2 e^{\Phi_0 '})]\cr
c &={1\over {\hat{\Delta}}}({{\hat q}_2}^2 e^{\Phi_0 '}- {{\hat p}_2}^2 e^{\Phi_0})\cr
d &={1\over {\hat{\Delta}}}[{{\hat p}_2}^2 (e^{\Phi_0 '}\chi_0 '+e^{\Phi_0}\chi_0)+
   e^{\Phi_0 '}({\hat p}_1 {\hat p}_2+{\hat q}_1 {\hat q}_2)],\cr
}}
being $\hat{\Delta}$ given by:
\eqn\dels{
\hat\Delta \equiv {\hat q}_2 {\hat p}_2 (e^{\Phi_0 '}\chi_0 '+e^{\Phi_0}\chi_0)+
 ({\hat q}_2 {\hat p}_1 e^{\Phi_0 '}+ {\hat q}_1 {\hat p}_2 e^{\Phi_0})
}
The constraint \conss\ now reads:
\eqn\consl{
e^{\Phi} ({\hat q}_1 + {\hat q}_2 \chi )^2+e^{-\Phi} {{\hat q}_2}^2 =
e^{\Phi '} ({\hat p}_1 + {\hat p}_2 \chi ')^2+e^{-\Phi '} {{\hat p}_2}^2.
}
The Killing spinors are computed
by using the equations \ks , \y , and 
\col .

Let us work out again an explicit example (\refs{\schw}). 
We will take a BPS fundamental string 
in the Einstein frame \refs\hor :
\eqn\fund{
\eqalign{
dS^2{_E}&=H(r)^{-{3\over 4}}(-dt^2 +dX^2_1)+H(r)^{1\over 4}
(dr^2+r^2 d\Omega_7)\cr
e^{-\Phi}&=\sqrt{H(r)},\,\,\,\,\,\,\,\,\,\,\, e^{-\Phi} * H^{(1)}=6 R^6 \epsilon_7,
 \,\,\,\,\,\,\,\,\,\,H^{(2)}=\chi =0,\cr
}}
being $H(r)$ the harmonic function:
\eqn\ard{
H(r)=1+{R^6\over r^6}.
}
We will take, moreover, the asymptotic conditions 
$\Phi_{0}'\equiv \Phi '(\infty )$
and $\chi_{0}'\equiv \chi '(\infty )$.

The equation \consl\ gives, for this case: 
\eqn\cnd{
{{\hat q}_1}^2=e^{\Phi_0 '} ({\hat p}_1+{\hat p}_2 \chi_0 ')^2+
e^{-\Phi_0 '} {{\hat p}_2}^2.
}
We can use this equation to eliminate ${\hat q}_1$ in equations \col\
and write \col\ in terms of ${\hat p}_1$ and ${\hat p}_2$ alone
(${\hat q}_1$ is given in \cnd\ in terms of ${\hat p}_1$ and ${\hat p}_2$) . 
This gives:
\eqn\rod{
\eqalign{
a &={{\hat p}_1\over {\hat q}_1}\cr
b &={1\over {\hat q}_1} [e^{\Phi_0 '}\chi_0 '({\hat p}_2 \chi_0 ' +{\hat p}_1)+ 
e^{-\Phi_0 '}{\hat p}_2] \cr
c &=-{{\hat p}_2\over {\hat q}_1}\cr
d &={e^{\Phi_0 '} \over {\hat q}_1} ({\hat p}_2 \chi_0 ' +{\hat p}_1).\cr
}}

Again, 
just plugging this background configuration
on \da\ and \h\ we get the $SL(2,Z)$ multiplet:
\eqn\muld{
\eqalign{
e^{\Phi '}&=c^2 H(r)^{{1\over 2}}+d^2 H(r)^{-1\over 2}\cr
\chi ' &={{ac H(r) +bd}\over {c^2 H(r)+d^2}}\cr
H^{(1)'} &=d H^{(1)}\cr
H^{(2)'} &=b H^{(1)} .\cr
}}

The coefficient $R$ of the harmonic form is now given, using \cnd, by: 

\eqn\tad{
{\hat q}_1 =\int{*S^{(1)}}=6 R^6 w_7\Rightarrow R_{({\hat p}_1, {\hat p}_2)}^6
\equiv R^6=
{\sqrt{{e^{\Phi_0 '} ({\hat p}_1 +{\hat p}_2 \chi_0 ')^2+e^{-\Phi_0 '} 
{{\hat p}_2}^2}}\over{6 w_7}},
}
where we denoted by $w_7$ the volume of the unit $S^7$. 
The equation \consl\ 
has to be interpretated, again, 
as equating the masses (more properly, tensions) of the components of 
the $SL(2,Z)$ multiplet when measured with the Einstein metric.

We can obtain the Killing spinor of the $({\hat p}_1,{\hat p}_2)$ strings from equation \ks :
\eqn\ku{
\epsilon_{({\hat p}_1 ,{\hat p}_2)}=
e^{{i\over 2}\Lambda_{({\hat p}_1,{\hat p}_2)}(r)}\epsilon_{F1},
}
where $\Lambda_{({\hat p}_1,{\hat p}_2)}(r)$ is given in this case by:
\eqn\lame{
\eqalign{
\sin \Lambda_{({\hat p}_1,{\hat p}_2)}(r)=
{{{\hat p}_2 [H(r)]^{1\over 2}}\over
\sqrt{e^{2 \Phi'_0} ({\hat p}_2 \chi'_0+{\hat p}_1)^2 + {\hat p}_2^2 H(r)}}\cr
\cos \Lambda_{({\hat p}_1,{\hat p}_2)}(r)=
{{e^{\Phi'_0} ({\hat p}_2 \chi'_0+{\hat p}_1)}\over
\sqrt{e^{2 \Phi'_0} ({\hat p}_2 \chi'_0+{\hat p}_1)^2 + {\hat p}_2^2 H(r)}},\cr
}}
and $\epsilon_{F1}$ is the Killing spinor of the fundamental $(1,0)$ string:
\eqn\killcf{
\epsilon_{F1}=
\left[ H(r)\right]^{-{3\over {16}}}
e^{-{\theta^7 \over 2} \Gamma_{\underline{\theta^7 r}}} 
\prod_{i=1}^6
\,\,e^{-{\theta^{i+1} \over 2} \Gamma_{\underline{\theta^i \theta^{i+1}}}}
\,\,\epsilon_0.
}
In the above equation, $\theta^i$ are the angles of the transverse $S^7$
and the product has to be taken from right to left. $\epsilon_0$ is a constant Weyl spinor satisfying:
\eqn\f{
\Gamma_{\underline {01}}\epsilon_0=-\epsilon_0^*
}

Now, we can compute the ADM masses of the electric multiplet 
in the modified Einstein frame similarly to our calculation
of the masses for the magnetic multiplets in the previous section. By comparing
\eqn\admd{
{\tilde g}_{00}\sim -1+{G_N\over {7 w_7}} {M\over r^6}.
}
with the asymptotic behaviour of the modified Einstein metric
associated to the metric in \fund\  
\eqn\oabd{
{\tilde g}_{00}\sim -1+{3\over 4} {{\sqrt{g_s^3} R_{\bar p}^6}\over r^6},
}
we obtain: 
\eqn\ewq{
M={21\over {24 {\alpha '}^4}}
\sqrt{({\hat p}_1 +{\hat p}_2\chi_0 ')^2+g_s^{-2}{{\hat p}_2}^2}.
}
Note that this ADM mass has units of mass per unit of brane volume ($1/(Lenght^2)$),
so it corresponds to the tensions of the $(p_1,p_2)$ strings. Note also
that the $p_1$ charged objects ``contribute" to the mass with terms of order
$p_1$ (as corresponds to the perturbative string modes) whereas
the $p_2$ charged objects ``contribute" to the mass with term of the order
$p_2/g_s$ (as it corresponds to D-brane solitons). Thus, the background \muld\
corresponds to bound states of fundamental strings and D1-branes
\refs{\witten}.

\newsec{Conclusions and Discussion.}

The equations of motion of type IIB supergravity exactly map the
equations of type IIB string theory through the identifications \id . The
S-duality symmetry of type IIB string theory corresponds to those
$U(1)$ local transformations of type IIB supergravity \u\ mantaining
the reality of the gauge \id . The $SL(2,Z)$ multiplets associated
to the solutions of type IIB string theory 
that we have considered here, are given in \da\ and \h,  
where the $SL(2,Z)$ constants are fixed by the expectation values
of the scalars and the electric or magnetic charges in \cons\ 
or \col . The Killing spinors of the $SL(2,Z)$ multiplets 
are then computed by \ks . 

In our approach, 
the self-duality of the D3 branes is reflected by the fact that the 
supergravity five-form is neutral under $U(1)$ transformations. Moreover,
the Killing spinors associated to D3 multiplets are related by global
rotations (see equations \y\ and \ks ), due to the constant
nature of the scalars for these solutions.

We have checked our general expressions with the
previously constructed $SL(2,Z)$ multiplets to which the solitonic
five brane (\refs{\lr}) and fundamental string (\refs{\schw}) belong,
obtaining the same result.
In \refs{\lr} and \refs{\schw} a different approach for these particular cases
was employed. In particular, the Eintein metric was not invariant under
the $SL(2,Z)$ transformation and the harmonic form coefficients 
\ta\ and \tad\ were found by a rescaling due to charge quantization arguments.
In other words, the solution to the effective IIB equations of motion
which was used to construct the multiplet was not itself in the multiplet.
In our approach, the harmonic form coefficient is fixed by
the consistency constraint \conss\ and the Einstein metric is invariant.
Moreover, we have provided the relation between the members of the multiplet in \cons\ and
\col .

Any $SL(2,Z)$ multiplet containing a given representative
with at least one non-vanishing charge (electric or
magnetic) associated to the NS-NS or R-R three forms can be 
straightforwardly constructed with our formalism. 
However, it does not cover exotic solutions as the D-Instantons 
(\refs{\gibb}) and
the D7-branes (\refs{\dsi}). These solutions are not charged with respect to
any of the three forms (or their duals) appearing in the theory. 
For the D-Instantons, one should adapt our formalism to the 
Euclidean space. The D7 case is a very subtle one.
This background solution is magnetically charged with respect to the
one-form R-R field strenght $d\chi$
(which, of course, is globally defined). This means that the 
field $\chi$ is not single-valued along one-cycles surrounding
the brane. We can, in this case, cover the cycle with two charts.
Inside any chart $\chi$ is single-valued, so we can
apply \dd\ safely. The  $SL(2,Z)$ transformations performed
in both charts are, however, not independent. We 
have to demand the equality of $d\Phi '$ and $d\chi '$ in
the overlapping regions (i.e., the field strenghts have to be globally defined).
It is not difficult from \dd\ to show that the $SL(2,Z)$ transformations
preserving the single-valued character of the one form field strenghts
associated to the potentials $\Phi$ and $\chi$ are those with $c=0$, 
which just change the magnetic charge associated to $d\chi$ and
the expectation values of the scalar fields.
The transformations with $c\neq 0$ lead to non single-valued field
strenghts. This suggests that bound states with D7 branes are absent
\refs{\witten}.

\bigskip
{\bf Acknowledgements}

We are grateful to J. Edelstein, A. V. Ramallo and 
J. S\'anchez Guill\'en for useful discussions and comments.
This work was supported in part by DGICYT under grant PB96-0960 
and by the European Union TMR grant ERBFMRXCT960012.

\listrefs
\end